\documentclass[final,3p,times,twocolumn,sort&compress]{elsarticle}

\usepackage[]{lineno}
\usepackage{subcaption}
\usepackage{graphicx}
\usepackage{overpic}
\usepackage[utf8]{inputenc}  
\usepackage[T1]{fontenc}
\usepackage{color}
\usepackage{amsmath}
\usepackage{amssymb}
\usepackage[separate-uncertainty=true]{siunitx}
\usepackage{dcolumn}
\usepackage{multirow}
\usepackage{array}
\usepackage{booktabs}
\usepackage[version=4]{mhchem}
\usepackage{tikz}
\usetikzlibrary {angles,arrows.meta,calc} 
\usepackage{tikz-imagelabels}
\usepackage{amsfonts}

\journal{Nuclear Instruments and Methods in Physics Research A}

\modulolinenumbers[5]
\newcolumntype{d}[1]{D{.}{.}{#1}}

\usepackage[bookmarks=true, pdfborder={0 0 0}, colorlinks=true, linkcolor=blue, citecolor=blue, filecolor=blue,urlcolor=blue]{hyperref}
\usepackage[capitalise]{cleveref}
\begin{document}


\title{Gas-jet target with online interferometric thickness measurement for nuclear astrophysics}

\author[HZDR,TUDD]{Anup Yadav}
\author[HZDR]{Daniel Bemmerer}\ead{d.bemmerer@hzdr.de}
\author[HZDR]{Fabian Donat}
\author[FTMC]{Juozas Dudutis}
\author[HZDR,TUDD]{Sören Göhler}
\author[HZDR]{Maik Görler}
\author[HZDR,TUDD]{Maxim Hilz}
\author[HZDR]{Arie Irman}
\author[FTMC]{Miglė Mackevičiūtė}
\author[HZDR]{Konrad Schmidt}\ead{konrad.schmidt@hzdr.de}
\author[HZDR]{Manfred Sobiella}
\author[FTMC]{Vidmantas Tomkus}
\author[TUDD]{Kai Zuber}

\address[HZDR]{Helmholtz-Zentrum Dresden-Rossendorf (HZDR), Bautzner Landstr. 400, 01328 Dresden, Germany}
\address[TUDD]{Technische Universität Dresden, Institut für Kern- und Teilchenphysik, Zellescher Weg 19, 01062 Dresden, Germany}
\address[FTMC]{Center for Physical Sciences and Technology (FTMC), Savanorių 231, 02300 Vilnius, Lithuania}

\date{\today}

\begin{frontmatter}

\begin{abstract}
A new jet gas target system has been developed for the Felsenkeller \qty{5}{\mega\volt} underground ion accelerator for nuclear astrophysics. It provides either a \qty{1.5e18}{atoms/\square\cm} thick cylindrical jet or a \qty{8e17}{atoms/\square\cm} thick wall of nitrogen gas, with a surface of \qtyproduct[product-units = power]{10x10}{\mm} to be seen by the ion beam.

The system includes a de Laval type nozzle and altogether five pumping stages: In addition to the jet catcher and the jet chamber surrounding it, there are three stages connecting the jet to the ion accelerator. Behind the jet chamber, as seen from the ion beam, a windowless static-type gas target and, subsequently, a beam calorimeter have been installed.

This work describes the offline tests of the gas target system prior to its installation on the beam line of the Felsenkeller accelerator. 

The thickness of the jet has been determined using three different methods: By computational fluid dynamics simulations, with a Mach-Zehnder interferometer, and by $\alpha$-energy loss using a mixed $\alpha$ source. The three methods were shown to be in agreement. For 0-6 bar inlet gas pressure, a linear relationship between inlet pressure and jet thickness has been found.  

Different shapes of de Laval type inlet nozzles, both circular and slit-type, have been manufactured from fused silica glass or stainless steel and tested using measurements and simulations. The power and stability of the beam calorimeter have been tested. The interferometry has been shown to work reliably and to give two-dimensional projections of the gas jet with sub-mm resolution. 
\end{abstract}


\end{frontmatter}

\section{Introduction}

Radiative capture reactions are important for energy production and nucleosynthesis in stars. Due to the repulsive Coulomb barrier, the reaction cross section at astrophysically relevant energies is very low, requiring stable, high target thickness, high beam intensity, and ultra-low background experiments, typically in an underground laboratory such as LUNA \cite{Broggini18-PPNP} or the Dresden Felsenkeller \cite{Bemmerer25-EPJA}. 
In this type of laboratories, frequently either solid \cite[e.g.]{Skowronski23-PRC} or static-type gaseous \cite[e.g.]{Ferraro18-PRL, Ferraro18-EPJA, Masha23-PRC, Masha25-EPJA} targets are being used. Gas cell targets with thin windows \cite{Toth23-PRC} have not yet been used underground, also due to limitations on beam current. 

Jet-type gas targets, instead, offer an attractive alternative,  allowing for cross-section measurements using both direct and indirect kinematics \cite[e.g.]{Schmidt18-NIMA}. They are well documented in the literature
\cite{Becker1954-ZfP, Bethge1976-ToNC, Tietsch1979-NIM, Kreisler1980-NIM, Bittner1979-NIM, Becker1982-NIM, Shapira1985-NIMA, Gruber1989-NIMA, Zapf1995-RSI, Schmidt1997-HyInt, Hosokai2002EPAC, Schmid2012-RoSI, Kontos2012-NIMA, Favela2015-JoPCS,Favela2015-PRSTaB, Rapagnani2017-NIMB, Rapagnani2023-NIMA, Chipps14-NIMA, Schmidt18-NIMA, Taeschner11-NIMA, Schlimme2021-NIMA, Grieser2018-NIMA}, and have been used also in other fields beyond nuclear astrophysics \cite[e.g.]{2022PhRvC.106d4610W}. The main characteristics of selected jet gas targets of relevance to the present work are summarized in Table \ref{tab:gas_targets}.

\begin{table*}[h]
    \centering
    \caption{Jet gas type, dimensions, thickness, and other characteristics of selected gas jet targets worldwide. Applications include nuclear astrophysics (Nucl. Astro.) and hadron physics (Hadron Phys.); techniques for thickness measurement include energy loss measurements ($\Delta E$), elastic scattering (ES), and laser interferometry (LI).}
    \label{tab:gas_targets}
    \begin{tabular}{l l c c c S[table-format=-1.3] c c}
    \toprule
    \multirow{2}{*}{Name} & \multirow{2}{*}{Location} & \multirow{2}{*}{Application} & \multirow{2}{*}{Gas} & Dimension & {Thickness} & {Thickness}    & \multirow{2}{*}{Ref.} \\
            &   &   &   & [mm] &  {[\qty{e18}{\per\square\cm}]} & {measured by} &   \\
    \midrule
    HIPPO   & Notre Dame    & Nucl. Astro.        & He        & $\varnothing=2.2$   & 0.267 & $\Delta E$(\ce{^4He}), ES & \cite{Kontos2012-NIMA}  \\
    \addlinespace
    ERNA    & Caserta       & Nucl. Astro.        & He        & cylindrical       & 1.97  & $\Delta E$(\ce{^12C})      & \cite{Rapagnani2017-NIMB} \\
    \addlinespace
    MAGIX   & Mainz         & Hadron Phys.    & \ce{H2}   & $\varnothing=2..3$   & 5     & LI                & \cite{Grieser2018-NIMA}  \\
    \addlinespace
    JENSA   & East Lansing  & Nucl. Astro.        & He        & $\varnothing=2.03$   & 9.0   & $\Delta E$(\ce{^4He})     & \cite{Schmidt18-NIMA} \\
    \addlinespace
    \hline
    \addlinespace
    Felsenkeller    & Dresden  & Nucl. Astro.  & N$_2$ & \numproduct{10x2.4} & 0.77    & $\Delta E$($^4$He), LI  & This work \\
    \bottomrule
    \end{tabular}
\end{table*}

There are some pitfalls, however, especially related to the total target thickness that must be carefully measured \cite{Kraewinkel1982-ZfPA, Filippone1986-ARNPS}. 

The present work reports on a newly developed gas-jet target system. It has been designed to meet the precision measurement demands of modern nuclear astrophysics \cite{Acharya24-RMP}. It follows the models established by LUNA at Gran Sasso, Italy, \cite{Ferraro18-EPJA,Masha25-EPJA} and by JENSA at SECAR, Michigan State University \cite{Schmidt18-NIMA}. In addition, a Mach-Zehnder type laser interferometer is used \cite{Couperus16-NIMA}. Its pumping scheme is designed to handle air, \ce{N2}, \ce{Ar}, \ce{He}, \ce{H2}, and \ce{CO2}, but due to time constraints, the target has only been commissioned for nitrogen gas so far, with the results reported in this work. 
The mechanical structure of the target allows the close placement of several large high-purity germanium (HPGe) detectors near the jet for low counting rate radiative capture experiments.

Summarizing, the salient characteristics of the new target system presented here are the possibility to use a wall-type jet with close to rectangular profile, the online interferometric thickness measurement during an in-beam irradiation, and the optimization for low-energy radiative capture reactions.

This work is structured as follows. The overall setup is described in \cref{sec:Description}, and the crucial gas nozzles in \cref{sec:Nozzles}. \cref{sec:laser-interferometry} discusses the Mach-Zehnder interferometer setup and results. The $\alpha$-energy loss measurements are described in \cref{sec:AlphaEnergyLoss}. \cref{sec:Discussion} offers a critical discussion of the measurements and simulations. A summary and outlook are given in \cref{sec:Outlook}.

\section{The gas target system}
\label{sec:Description}

\begin{figure}[htb]
    \centering
    \includegraphics[width=\columnwidth, clip,trim=0cm 1cm 0cm 0cm]{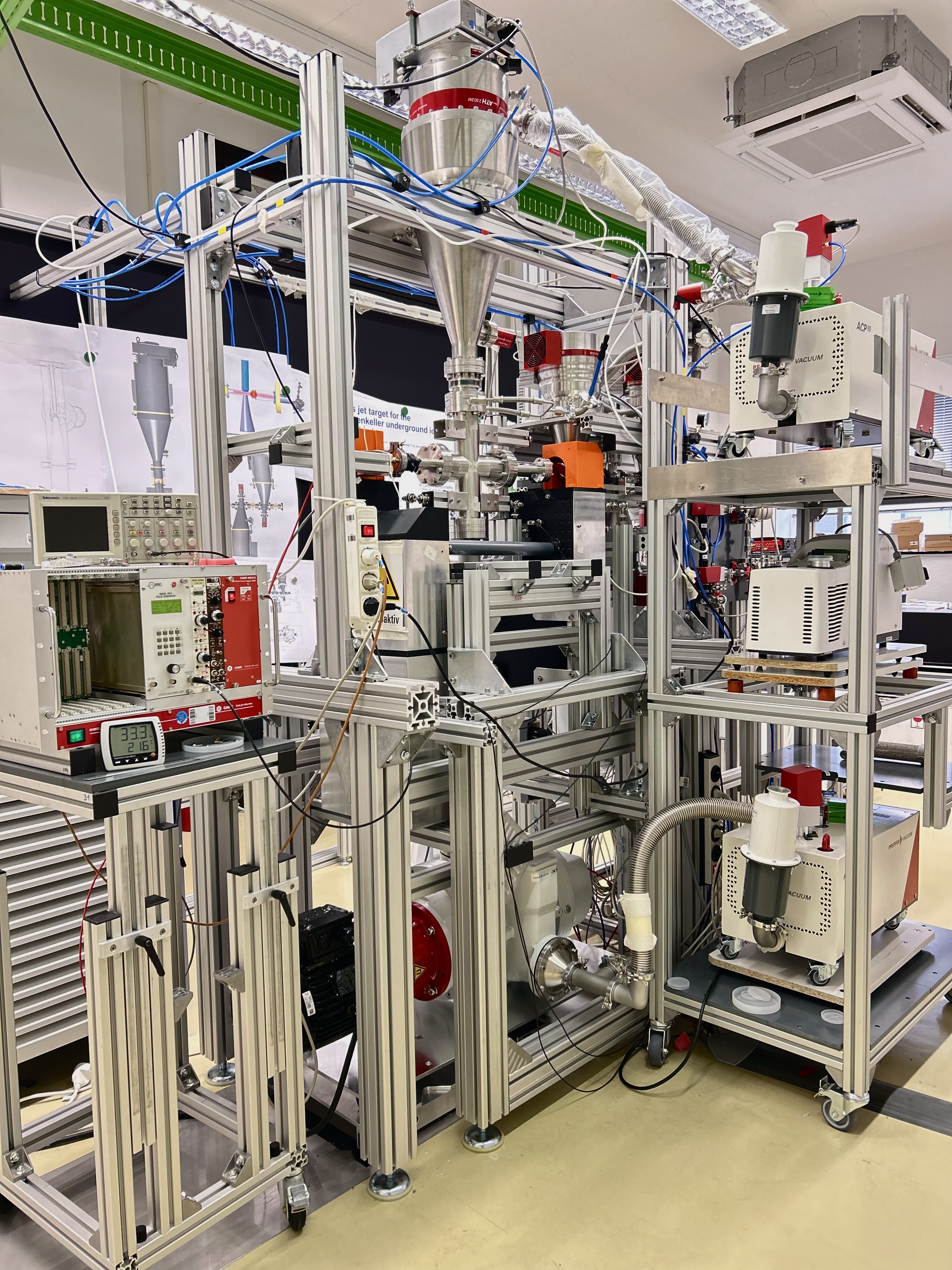}
    \caption{Photograph of the gas target system. Downward looking on top is the ATH~2303~M turbopump (chamber). On the bottom, partially hidden by Bosch profiles, the Okta~1000~M (Catcher). The structure on the right hosts the ACP backing pumps. The laser and optics are in two black boxes on either side of the center. See text for details.}
    \label{fig:Photo}
\end{figure}

The main components of the system (\cref{fig:PumpingScheme} and \cref{fig:Photo}) are described below. They include the jet chamber and catcher (\cref{subsec:Catcher}), the pumping scheme (\cref{subsec:PumpingScheme}), the resulting pressure profile (\cref{subsec:PressureProfile}), and the beam calorimeter for the measurement of the beam intensity (\cref{subsec:BeamCalorimeter}). The gas nozzles merit special attention and are discussed in a separate \cref{sec:Nozzles}.

\subsection{Jet chamber and catcher}
\label{subsec:Catcher}

The incoming gas (typical chemical purity of \qty{>99.999}{\percent}) is transported through stainless steel pipes to a downward facing de Laval type nozzle with a throat of \qty{0.8}{\square\mm}.
The gas jet then expands through the nozzle. The gas particles achieve high velocities at the nozzle's exhaust, typically higher than the speed of sound in the gas. Further details can be found below, \cref{sec:Nozzles}.

Most of the gas inside the supersonic jet is subsequently collected by a cone-shaped catcher (\cref{fig:PumpingScheme}). 
The distance between the nozzle and catcher is  \qty{10}{\mm}, a compromise between optimal capture of the jet and giving enough space for the 5\, mm diameter ion beam. 
The catcher has a smooth surface and a sharp-edged opening, designed to facilitate non-turbulent gas flow. The gas is then pumped out by a magnetically coupled universal booster pump backed by multi-stage roots pumps. The fraction of the gas that does not enter the catcher is removed by the subsequent pumping stages (\cref{subsec:PumpingScheme}).

The total width of the jet chamber is \qty{6.3}{\cm}. 
The narrow design allows for relatively close geometry of $\gamma$-ray detectors around the jet chamber.
For example, a sevenfold Euroball-type Cluster detector can be placed at a distance between detector endcap and gas jet of \qty{3.9}{cm}, if the BGO escape-suppression shield is omitted. Slightly closer geometries are possible for single germanium or sodium iodide detectors.

\subsection{Pumping scheme}
\label{subsec:PumpingScheme}

\begin{figure*}[htb]
    \centering
    \begin{overpic}[width=\textwidth, clip, trim=1.3cm 2cm 0.8cm 2cm]{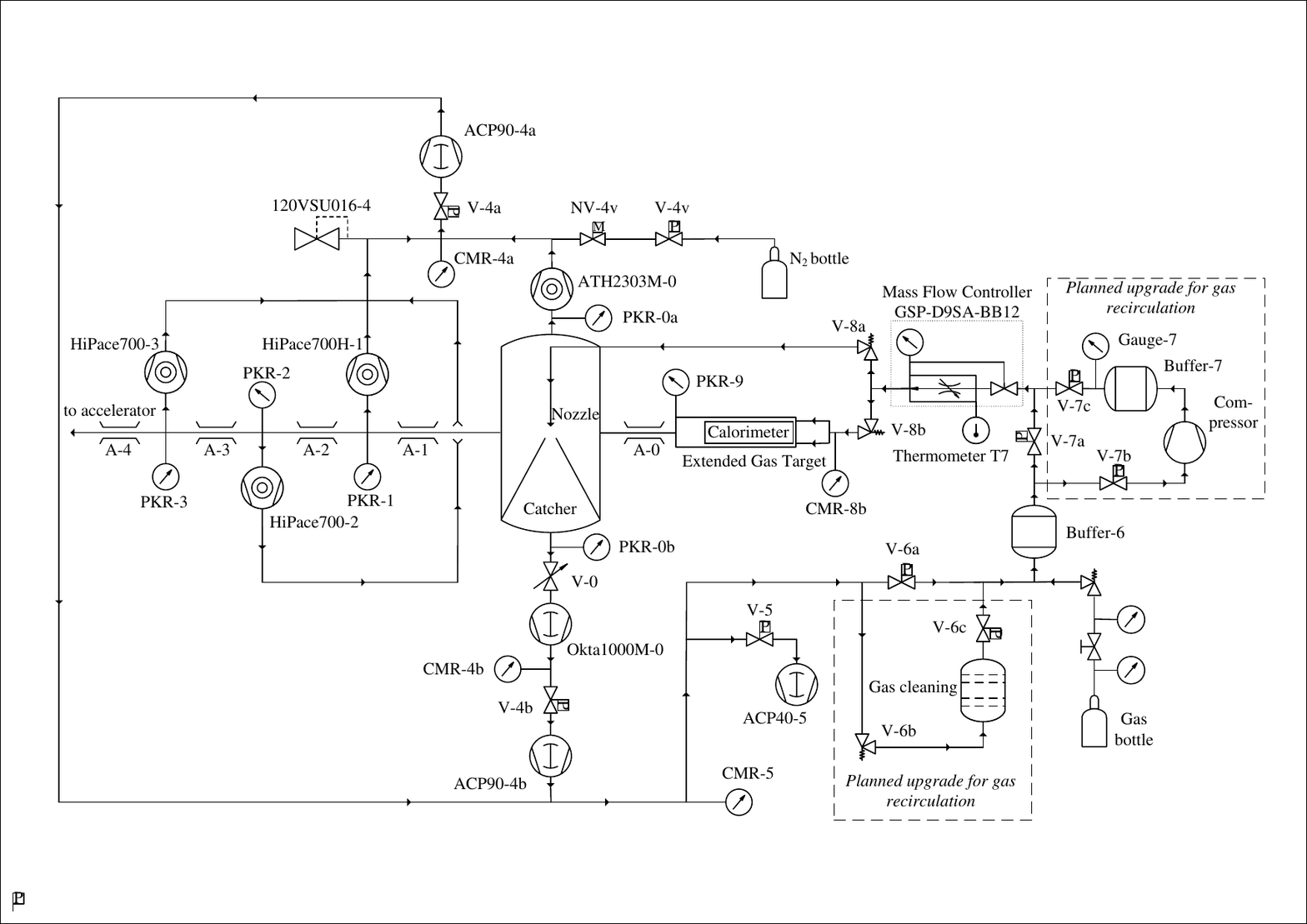}
        \put(54,57){
            \small
            \begin{tabular}{|c c c c c c|}
                \hline
                Aperture       & A-4   & A-3   & A-2   & A-1   & A-0\\
                \hline
                Diameter (mm)  & 5.1   & 5.1   & 5.1   & 6.0   & 5.1   \\
                Length (mm)    & 60   & 60 & 60    &  184.75    & 180   \\
                \hline
            \end{tabular}
        }
    \end{overpic}
    \caption{Schematic view of the Felsenkeller gas target. Gas flow directions are indicated by arrows. The dimensions of the apertures A-0 to A-4 are given in the upper right corner. 
    Pumps shown are of turbomolecular (ATH, HiPace), Roots (Okta), and multi-stage Roots (ACP) types,  vacuum gauges of full range (PKR) or capacitance (CMR) types, gate valves (V), and needle dosing valves (NV). 
    Two planned future upgrades are shown in dashed boxes.}
    \label{fig:PumpingScheme}
\end{figure*}

The pumping scheme is arranged in altogether five stages, which are described below, in turn. Typical pressure values are given for 4\,bar nitrogen as inlet gas. Further improvements of the pumping scheme to minimize the residual gas and the study of the pumping efficiency of other gases are not part of this work and will be reported elsewhere.

The apertures A-0 to A-4 described below separate the stages from each other. They are all cylindrical and made out of copper, with the inner diameter $d$ and length $l$ given.  Flexible bellows are installed on both sides of each aperture to dampen vibrations originating from the pumps.

\begin{figure}[htb]
    \centering
    \includegraphics[width=\columnwidth,,clip,trim=0cm 1cm 2cm 2.5cm]{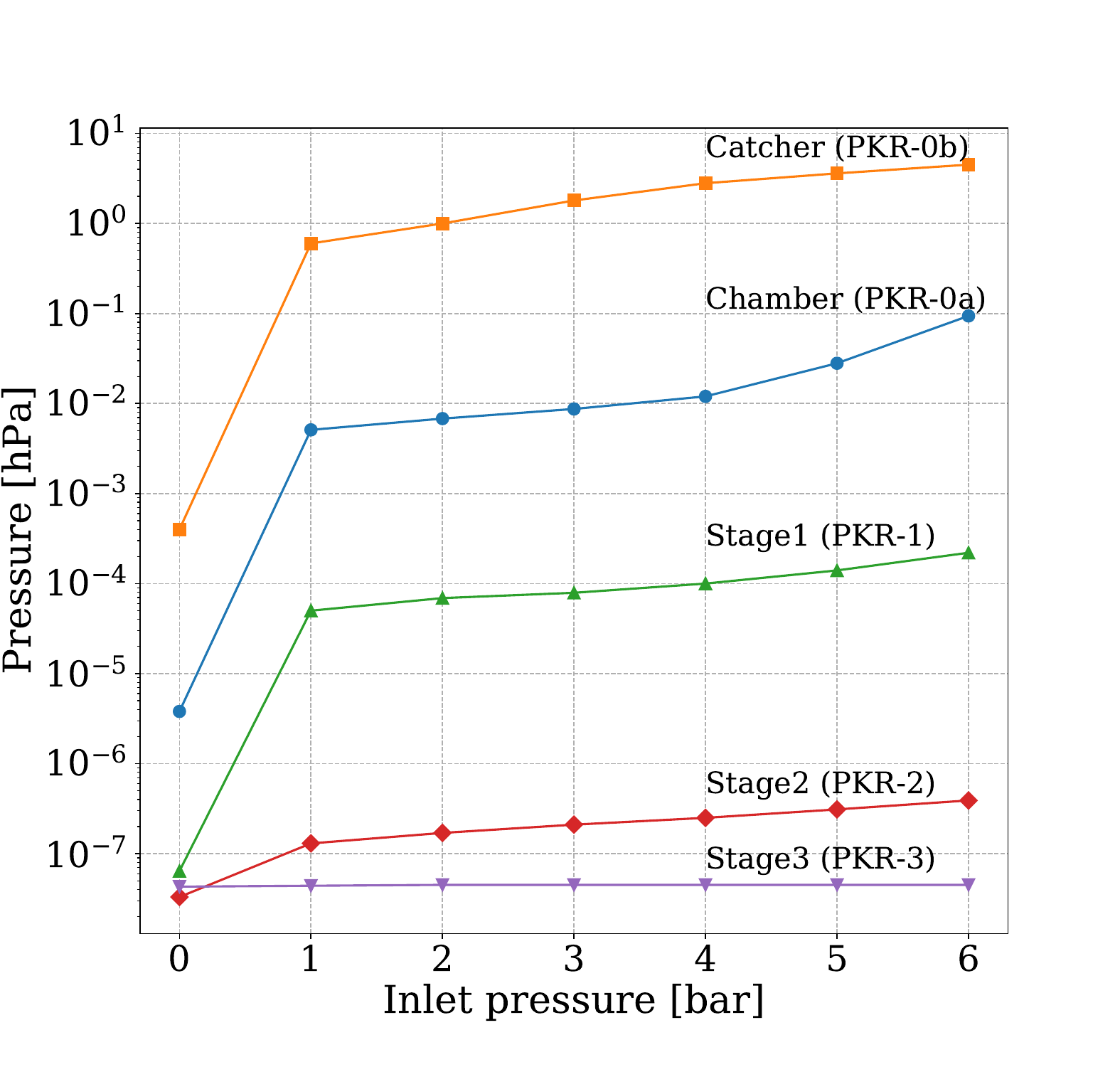}
    \caption{Pressure profile as a function of nozzle inlet pressure for nitrogen gas, cylindrical nozzle CF-64-1, and cone-shaped catcher C20 of 20\,mm diameter.}
    \label{fig:PressureProfile}
\end{figure}

{\bf Catcher:} At a distance of 10\,mm below the lower end of the gas nozzle, a circular catcher of 20\,mm diameter is arranged. Here, the typical pressure is 3 hPa (\cref{fig:PressureProfile}). The catcher collects >99\% of the gas included in the expanding jet. It is evacuated by a Roots pump (Pfeiffer Okta 1000M), backed by a multi-stage Roots pump (Pfeiffer ACP90). 

{\bf Chamber:} The chamber surrounding the gas catcher receives gas from the jet that is not collected or that is reflected back from one of the catcher walls (backstreaming). It has a typical pressure of \qty{1.3e-2}{hPa} and is evacuated by a Pfeiffer ATH 2303 M turbomolecular pump (2150 l/s nominal speed). 

{\bf Stage 1:} Separated from the chamber by aperture A-1 ($d$ = 6.0\,mm, $l$ = 185\,mm), the first differential pumping stage has a typical pressure of \qty{1.0e-4}{hPa} and is evacuated by a Pfeiffer HiPace700H turbomolecular pump (685 l/s nominal speed).

{\bf Stages 2 and 3:} The second and third pumping stages are both separated from the respective preceding pumping stage by an aperture (A-2 or A-3, respectively) with $d$ = 5.1\,mm, $l$ = 60\,mm. Each of them is evacuated by a Pfeiffer HiPace700 turbomolecular pump (685 l/s nominal speed). Stages 2 and 3 show typical pressures of \qty{3e-7}{hPa} and \qty{5e-8}{hPa}, respectively. 

After Stage 3, another aperture (A-4) with equal dimensions as A-3 and A-2 is separating the gas target system from the accelerator and beam line. 

{\bf Static-type gas target:} After the jet chamber (to the right in \cref{fig:PumpingScheme}), a static-type windowless gas target chamber is positioned. This chamber is prepared for future use and is not characterized in the present work. It includes a beam calorimeter for the calorimetric measurement of the ion beam intensity, and it is evacuated through the above mentioned chamber stage.

The turbomolecular pumps in stages 1-3 and the chamber are together backed by a Pfeiffer ACP90 multi-stage oil-free Roots pump that releases its exhaust into the room. In order to accommodate high gas loads, all turbomolecular pumps in the system are watercooled.

\subsection{Pressure profile}
\label{subsec:PressureProfile}

Using nitrogen as inlet gas, the cylindrical nozzle CF-64-1, and the cone-shaped catcher C20 of 20\,mm diameter, the pressure has been measured in each of the pumping stages (\cref{fig:PressureProfile}), using the pressure gauges shown in \cref{fig:PumpingScheme}. 

Without inlet gas (0 bar inlet pressure), pumping stages 1-3 show pressure values in the \qty{e-8}{hPa} range (\cref{fig:PressureProfile}). When the shutter valve V-0 (\cref{fig:PumpingScheme}) is open, the catcher pressure is \qty{4e-4}{hPa}, close to the optimum achievable for the Okta 1000M pump without load and the chamber pressure is \qty{3.8e-6}{hPa}, indicating some gas load from possible backstreaming from the Okta. Both values decrease significantly 
when valve V-0 is closed.

For \qtyrange[range-units=single]{1}{4}{bar} inlet has pressure, there is an approximately constant pressure ratio of \qty{5e6}{} : \qty{12e3}{} : \qty{120}{} : 1 for Stage 2 : Stage 1 : Chamber : Catcher. As a result, the pressure in Stages 1 and 2 is so low that the energy loss of the incident ion beam in the residual gas can be neglected comparable to the energy loss in the jet itself. At the highest inlet pressure, \qty{6}{bar}, the integrated thickness of the residual gas is $\sim$4\% of the jet thickness, demanding a small correction of the beam energy.

For \qtyrange[range-units=single]{5}{6}{bar} bar inlet pressure, the chamber pressure rises more quickly than the catcher pressure. This is due to the fact that the pumping speed of the large, watercooled turbomolecular pump ATH2303M starts to quickly drop from its maximum of 2150 l/s for pressures above \qty{1e-2}{hPa}. This drop is due to the fact that the ATH is not designed for pressures higher than \qty{1e-2}{hPa}, and it leads to an upper limit of 6 bar inlet pressure. 
This may in the future be overcome either by an improved catcher geometry, or by improved backing capacity for the Okta1000M evacuating the catcher. 

At all inlet pressure values studied, the Stage 3 pressure is stable at \qty{4.5e-8}{hPa}, independent of inlet pressure. This shows that there is no parasitic gas flow to the accelerator beam line from the gas target system.

\subsection{Beam calorimeter}
\label{subsec:BeamCalorimeter}

\begin{figure*}[t]
    \centering
    \includegraphics[width=\textwidth,,clip,trim=0cm 0cm 0cm 0cm]{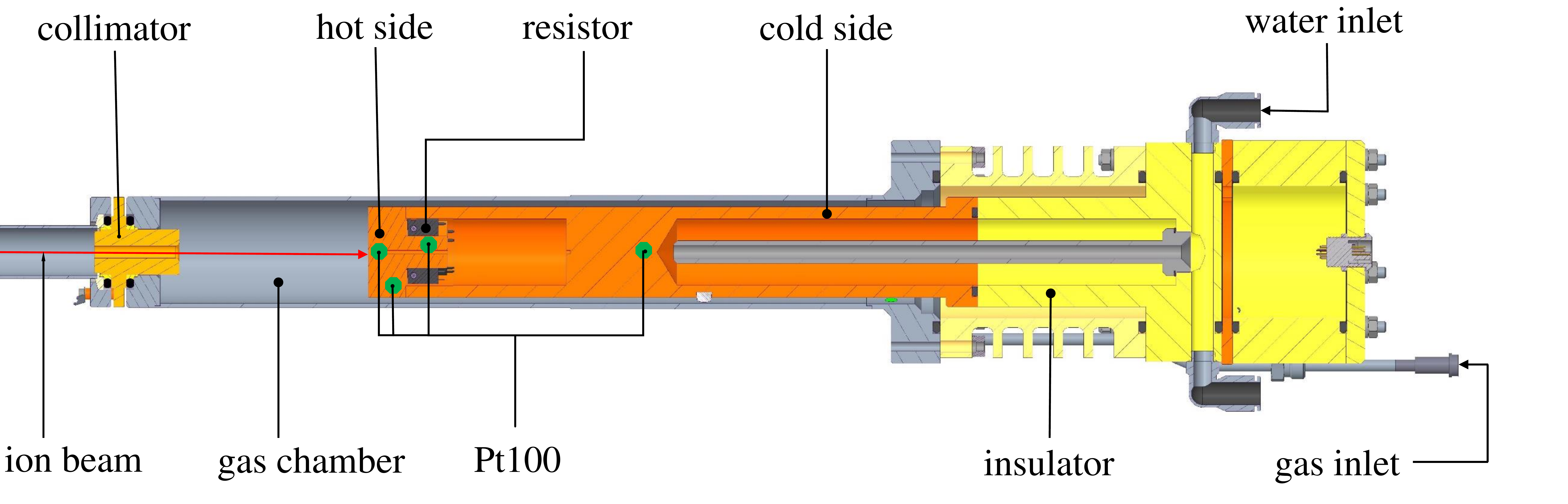}
    \caption{Schematic drawing of the calorimeter inside the static-type gas target. The beam hits the calorimeter from the left side.}
    \label{fig:CalorimeterScheme}
\end{figure*}

In order to measure the current in gaseous environments such as a gas target system, due to the prevalence of secondary electrons the usual electric current integration cannot be used. Instead, the beam intensity is measured by a power compensation calorimeter here. 

The design is based on the LUNA beam calorimeter \cite{Casella02-NIMA,Ferraro18-EPJA}. The main calorimeter body is made of copper, operating on the principle of constant temperature gradient between the hot and cold sides of the device.

The cold side is maintained at a low temperature (typically 1\,$^\circ$C) by circulating a coolant (deionized water, 15\% glycol admixture, biocide added) through a chiller with 600\,W nominal cooling power (Julabo 600F). The hot side is kept at constant high temperature using eight resistors, with nominally \qty{2}{\ohm} each, and by the incident ion beam. The more power the beam provides, less power is required from the heating resistors to keep the hot side at a constant temperature. The current passed through the resistors is controlled by a feedback-controlled circuit managed by a LabVIEW \cite{Labview2006} application, ensuring a consistent temperature gradient between hot and cold sides. 

The temperatures are measured by four Pt100 thermoresistors placed at various places inside the calorimeter (\cref{fig:CalorimeterScheme}). The electrical power is calculated from the measured current and voltage for the heating resistor chain. 

\begin{figure}
    \centering
    \includegraphics[width=\columnwidth, clip, trim={0.8cm 1.3cm 0cm 2.7cm},clip]{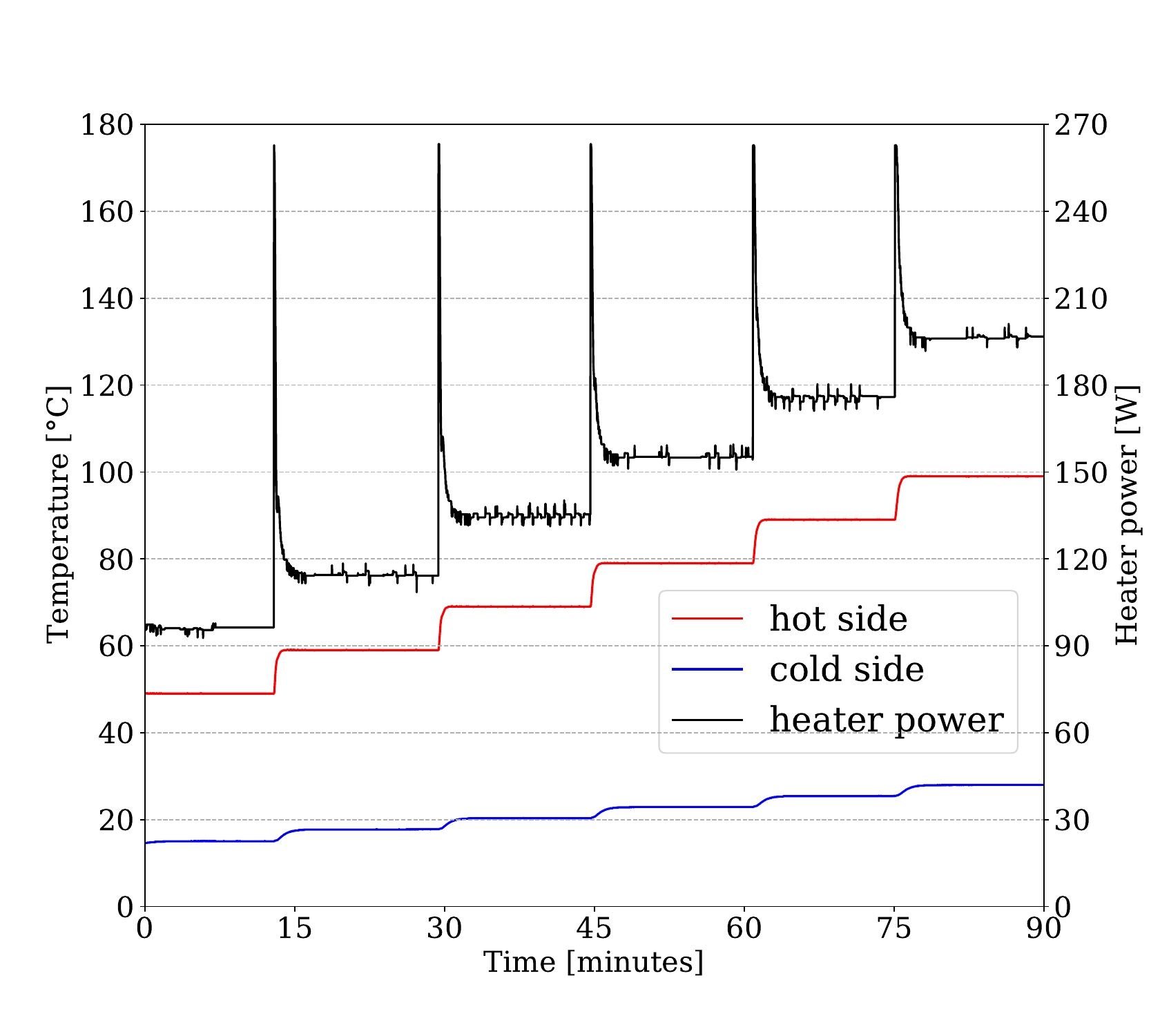}
    \caption{Power tests of the beam calorimeter at hot side temperatures 50-100\,$^\circ$C and fixed chiller temperature of 1\,$^\circ$C.}
    \label{fig:CalorimeterPower}
\end{figure}

The results of power tests with different hot side preset temperatures from 50-100\,$^\circ$C and chiller kept constant at 1\,$^\circ$C are shown in \cref{fig:CalorimeterPower}. It is found that the power and the temperatures of hot and cold sides typically stabilize less than 5 minutes after a change in hot side preset temperature. The highest observed power is 200\,W, at a hot side preset temperature of 100\,$^\circ$C. For long-term operation, 70\,$^\circ$C hot side preset temperature is adopted, corresponding to 134\,W in total calorimeter power and a cold side temperature of 20\,$^\circ$C.

The electrical calibration of the calorimeter will be described in a follow-up publication when the setup is mounted on the beam line. 

\section{Nozzle design and simulation}
\label{sec:Nozzles}

In this section, the different nozzles tested are described (\cref{subsec:NozzlesTested}), and the gas flow is simulated by the ANSYS Fluent computational fluid dynamics code (\cref{subsec:NozzleSimulations}).

\subsection{Nozzles tested}
\label{subsec:NozzlesTested}

\begin{figure}[htb]
    \centering
    \includegraphics[width=\columnwidth,clip,trim=4.8cm 12.5cm 5.3cm 5cm]{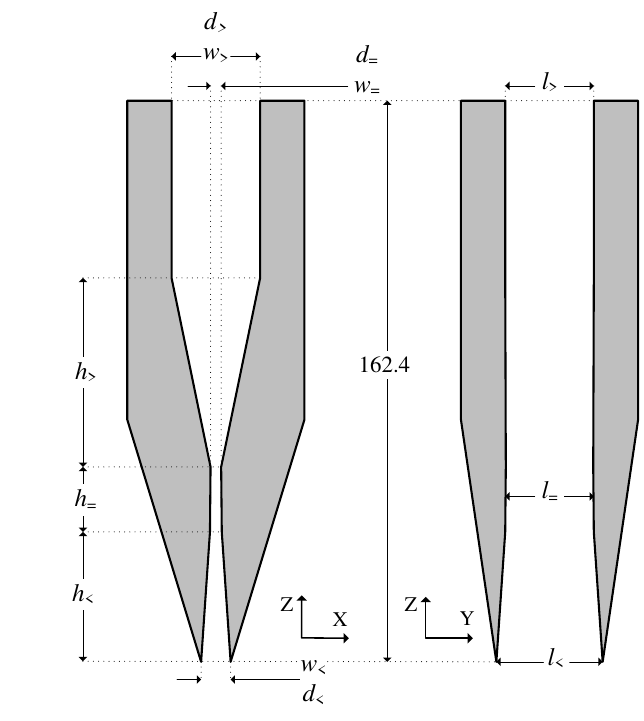}
    \caption{Schematic view of the gas nozzles tested, gas flowing from top to bottom. Left panel, side view (cylindrical and slit nozzles). Right panel, front view (slit nozzles only). See \cref{tab:NozzleDimensions} for dimensions.}
    \label{fig:Nozzles}
\end{figure}

Two different general shapes of de Laval-type nozzles were used: cylindrical and slit-type.  Cylindrical nozzles were used to test general parameters and to validate the simulation, which assumes cylindrical symmetry. Slit type nozzles were also used, because they are closer to the final experimental configuration. 

The total length of each nozzle tested here is \qty{162.4}{\mm}. This total length includes a \qty{40}{\mm} long de Laval nozzle (either from fused silica or from stainless steel) and a \qty{122.4}{\mm} long nozzle adapter (stainless steel). 

The fused silica nozzles were fabricated using the nanosecond laser rear-side milling technique \cite{Tomkus2024-JPP} at the Center for Physical Sciences and Technology (FTMC) in Vilnius, Lithuania, resulting in large aspect ratio, fast processing and smooth surfaces with an average roughness of \qty{1.4}{\um} \cite{Tomkus2019-ASS}. The stainless steel nozzles were produced by electrical erosion with a \qty{0.5}{mm} thin wire resulting in a roughness of \qty{3.2}{\um} according to the manufacturer.

In order to optimize the nozzle material and dimension, altogether four fused silica and five stainless steel nozzles have been tested (\cref{tab:NozzleDimensions}). In addition to their cylindrical (C) or slite-type (S) geometry and the material and producer (F for FTMC glass, H for HZDR steel), these nozzles differed mainly in the dimensions of the converging sections (height $h_>$=11-151 mm), and of the crucial throat (height $h_=$=0-4 mm). 

In addition to the nozzles, several gas catchers of different shapes and sizes were also tested. In particular, it was attempted to match circular nozzles with circular catchers and slit-type nozzles with rectangular catchers. Catchers tested included five cone-shaped catchers with diameters of \qtylist[list-units = single,list-final-separator = {, and }]{15;18;20;22;25}{\mm}, as well as five rectangular catchers of \numproduct{20x7}, \numproduct{20x10}, \numproduct{20x17}, \numproduct{25x10} and \numproduct{25x17} \unit{\square\mm} for the slit-type nozzles. 

Figure \ref{fig:Nozzles} shows the schematic design of the nozzles tested here. After a constant diameter inlet pipe, they include a converging section, the throat, and a diverging section, with dimensions given in \cref{tab:NozzleDimensions}. In the converging section of cylindrical nozzles, the inner diameter changes from $d_>$ over the convergent height $h_>$ to the throat diameter $d_=$. This remains in the throat (height $h_=$) and then changes over the divergent height $h_<$ to the convergent diameter $d_<$ at the outlet of the nozzle. For slit-type nozzles, instead of the diameters $d$ the widths $w$ and lengths $l$ are shown (\cref{fig:Nozzles}).


\begin{table}
    \centering
    \caption{Nozzle dimensions in mm. First, cylindrical nozzles (coded "C", diameter $d$ given). Afterwards, slit-type nozzles (coded "S", slit width $w$ and length $l$ given). The height of the diverging section is always $h_<$ = 10\,mm. See text and Figure \ref{fig:Nozzles} for details. }
    \label{tab:NozzleDimensions}
    \resizebox{\columnwidth}{!}{%
    \begin{tabular}{c*{8}{d{1}}c}
        \hline
        \hline
        Nozzle  & \multicolumn{3}{c}{Converging} & \multicolumn{3}{c}{Throat}   & \multicolumn{2}{c}{Diverging}  & Material\\ \cline{2-4}   \cline{8-9}
          & \multicolumn{1}{c}{$h_>$} &  & \multicolumn{1}{c}{$d_>$}   & \multicolumn{1}{c}{$h_=$}  &  & \multicolumn{1}{c}{$d_=$}     &   & \multicolumn{1}{c}{$d_<$}   & \\
        \hline
        CF-62-2    &151.4  &    &8       &1.0    &    &1.0      &    &2.8    &    glass\\
        CF-64-1    &25.5   &    &9.9    &1.0    &    &1.0      &    &2.7    & glass\\
        CH-65-1    &15.0   &    &10     &0.0    &    &1.0      &    &2.7    & steel\\
        CH-65-2    &14.0   &    &10     &1.0    &    &1.0      &    &2.7    & steel\\
        CH-65-3    &13.5   &    &10     &1.5    &    &1.0      &    &2.7    & steel\\
        CH-65-4    &13.0   &    &10     &2.0    &    &1.0      &    &2.7    & steel\\
        CH-65-5    &11.0   &    &10     &4.0    &    &1.0      &    &2.7    & steel\\ \hline
        ~\\
        Nozzle  & \multicolumn{3}{c}{Converging} & \multicolumn{3}{c}{Throat}   & \multicolumn{2}{c}{Diverging}  & Material\\ \cline{2-4}   \cline{8-10}
        & \multicolumn{1}{c}{$h_>$}  &\multicolumn{1}{c}{$w_>$}  & \multicolumn{1}{c}{$l_>$}   & \multicolumn{1}{c}{$h_=$}  &\multicolumn{1}{c}{$w_=$}  & \multicolumn{1}{c}{$l_=$}    & \multicolumn{1}{c}{$w_<$}  & \multicolumn{1}{c}{$l_<$}  & \\ \hline
        SF-61-2    &151.4  &8      &8.0    &1.0    &0.1    &8.0     &1.9    &9.8 & glass\\
        SF-63-2    &22.4   &8      &8.0    &1.0    &0.1    &8.0     &1.9    &9.8 & glass\\
        \hline
        \hline
    \end{tabular}
    }
\end{table}

The slowly converging nozzles provided us with a higher thickness of the jet. Our nozzle development process involved comprehensive computational fluid dynamics simulations to optimize the nozzle geometry.

\subsection{ANSYS computational fluid dynamics simulations}
\label{subsec:NozzleSimulations}

\begin{figure}[t!!]
    \centering
    \includegraphics[width=\columnwidth]{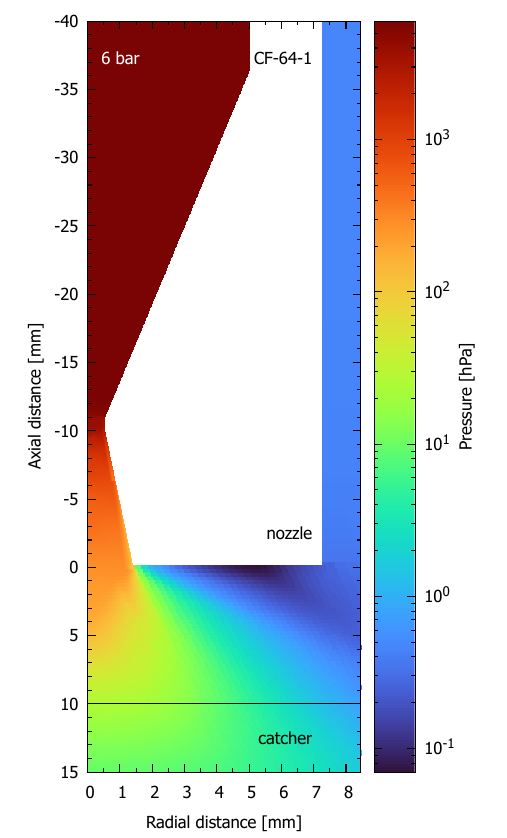}
    \caption{Simulated pressure profile of a nitrogen gas jet expanding in the jet chamber, for 6 bar inlet pressure and \qty{6.4e-2}{hPa} remaining gas pressure in the catcher. Nozzle CF-64-1 is shown in white. The catcher (not simulated) starts at 10 mm axial distance.
    }
    \label{fig:NozzleSimulation}
\end{figure}

In order to support the nozzle development process, computational fluid dynamic (CFD) simulations were performed using ANSYS Fluent (version~22.2) \cite{fluent2022ansys}, for nitrogen gas, assumed to behave like an ideal gas. The simulations utilized a density-based solver imposing two-dimensional axial symmetry. The SST k-omega viscous model and the energy equation were assumed. 

The simulations focused on modeling the gas jet within the nozzle and its expansion into the open vacuum chamber \cite{Meisel16-NIMA}. In order to limit the simulation effort, components such as catchers and apertures on both sides of the jet chamber were not modeled in details but instead taken into account as so-called ambient background, set to the measured pressure value in that area.
The mesh structure of the numerical solver included \num{107500} quadrilateral cells, with a concentration of meshes inside the nozzle and below it, where the jet expands. 

 Design parameters varied within the simulations include the nozzle dimensions $d_>$, $h_>$, $d_=$, $h_=$, $d_<$, and $h_<$ (\cref{fig:Nozzles}), the nozzle inlet pressure (1-6 bar), and the ambient pressure.   

Using the Abel transformation \cite{Pretzier1991}, the simulated density in units of \unit{\kg/\cubic\m} is then transformed (projected) to an areal thickness in units of \unit{atoms/\square\cm} (\cref{fig:NozzleSimulation}). The Figure shows how the jet forms in the first few mm below the exhaust (at 0 mm) of nozzle CF-64-1 and then dissipates at about 7 mm, slightly above the gas catcher which is at \qty{10}{mm} distance to the nozzle. 

The simulations have been repeated for nine different nozzles, and the average of the areal thickness in the area of the ion beam, taken to be a circle of \qty{5}{mm} diameter, centered \qty{5}{mm} below the nozzle, is listed in \cref{tab:JetCharacteristics}.

\section{Laser interferometry setup and methods\label{sec:laser-interferometry}}

\subsection{Experimental setup}

\begin{figure*}
    \centering
    \includegraphics[width=\textwidth,clip, trim=0.5cm 9cm 0.5cm 2.5cm]{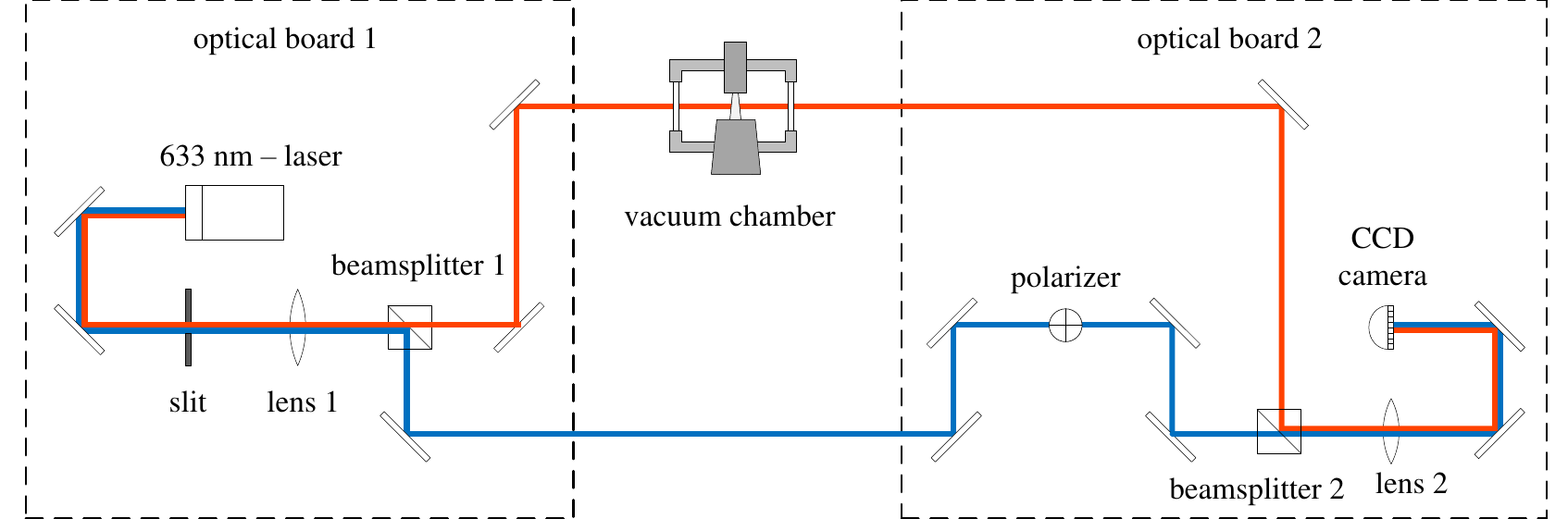}
    \caption{Schematic of the Mach-Zehnder interferometer used for {\it in situ} characterization of the gas jet. The two arms of the laser beam are shown in red (probe arm) and blue (reference arm), respectively. Mirrors are denoted by $\pm$45$^\circ$ inclined disks. The drawing is not to scale. See text for details.}
    \label{fig:InterferometryScheme}
\end{figure*}

For the jet density profile monitoring a Mach-Zehnder interferometer is used (\cref{fig:InterferometryScheme}). It uses a \qty{632.8}{\nm} helium neon (HeNe) laser with a power below \qty{5}{mW}, class 3R. Further components include two beam splitters, two lenses, twelve mirrors, and a linear polarizer. 

Initially, the laser beam passes through a \qty{0.5}{\mm} small pinhole to ensure spatial resolution and is then expanded to illuminate the entire gas jet with a maximum laser diameter of \qty{16}{\mm} using a lens. Subsequently, the laser beam is split using a 50:50 beam splitter into two arms called probe and reference arm, respectively.

The probe arm traverses the jet chamber and the jet itself, while the reference arm passes outside. Eventually, both arms are combined using another beam splitter and pass through a final lens and two mirrors. The lengths of both interferometer arms are approximately equal with a length of \qty{256}{\cm} each, to remain their difference within the coherence length (approximately 30 cm) of the laser. 

Due to the difference in optical path lengths between the two arms, there is an interference pattern. The pattern is recorded by a CCD camera placed in the plane of the image of the gas jet. The camera has an active area of 7.2$\times$5.4\,mm$^2$ (1600$\times$1200 monochromatic pixels with 12 bit depth) and is set to \qty{25}{\us} exposure time, recording 60 frames per second (for a typical frame recorded, see \cref{fig:Interferogram} below).

The typical observed path length differences are well below one wavelength, so it is enough to consider the phase difference between the two waves. In order to convert the observed phase difference to a gas thickness, an index of refraction of $n$ = 1.0002903 is used for nitrogen gas at 23$^\circ$C ambient temperature, 632.8 nm wavelength light. Furthermore, the reconstruction assumes that the reduced index of refraction $n-1$ depends linearly on the gas density. This is indeed true in the pressure range relevant for the present jet (0-200 hPa). There, deviations from linearity between pressure $p$ and reduced index of refraction $n(p)-1$ are below 10$^{-8}$, negligible for the present purposes \cite{Rumble2024}.

\subsection{Vibrations and their mitigation}

\begin{figure}[b!]
    \centering
    \includegraphics[width=\columnwidth, trim={8mm 13mm 12mm 17mm},clip]{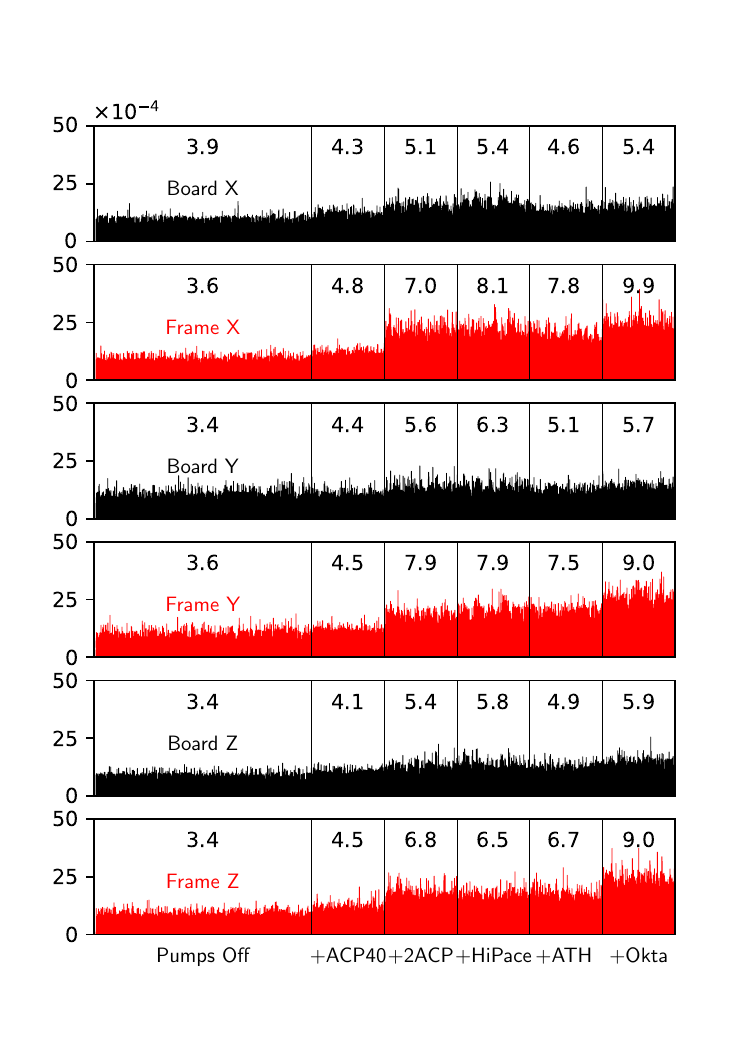}
    \caption{Vibration amplitude in multiples of the gravitational acceleration as a function of time, as pumps are switched on one after the other. The operating regime is listed below the time axis. Numbers given are the root-mean-square for each regime.}
    \label{fig:Vibrations}
\end{figure}

Even though the optical boards are completely separated from the structures holding the pumps and the vacuum chamber, some residual vibrations may disturb the interferometric measurement. 

In order to quantitatively study these vibrations, we utilized two KS943B100 triaxial piezoelectric accelerometers \cite{VibSensors}, defining the axes as follows: the X-axis along the beam line, the Z-axis vertically aligned with the gas-target system, and the Y-axis perpendicular to them. 

These sensors were placed on the frame holding the optical boards and on the optical boards themselves, after an insulating layer. It was found that for 6\,bar inlet gas pressure with a nitrogen jet, the root-mean-square vibration amplitude rose from a pumps-off baseline of (3.4-3.9)$\times$10$^{-4}$\,g to (5.4-5.9)$\times$10$^{-4}$\,g with all pumps running (\cref{fig:Vibrations}). 

Subsequently, several different additional damping regimes in particular for the large Okta pump were studied. For this case, it was found that hard rubber feet with an aluminum core provided better insulation for vibrations traveling through the concrete ground than no feet, or purely aluminum or rubber feet. In addition, bellows were introduced on the beam line to further mitigate vibrations traveling through the vacuum chamber. Sound waves were absorbed by adding a soundproof housing of the laser boards. 

After these improvements had been implemented, vibrations of (4.5-5.0)$\times$10$^{-4}$\,g remain in the x and y directions most relevant for interferometry, 20\% above the baseline level observed with all pumps off.

\subsection{Interferometer acquisition and processing\label{sec:data_processing}}

\begin{figure}[tb]
    \centering
    \includegraphics[width=.9\columnwidth,clip, trim=1cm 4.3cm 1.8cm 6cm]{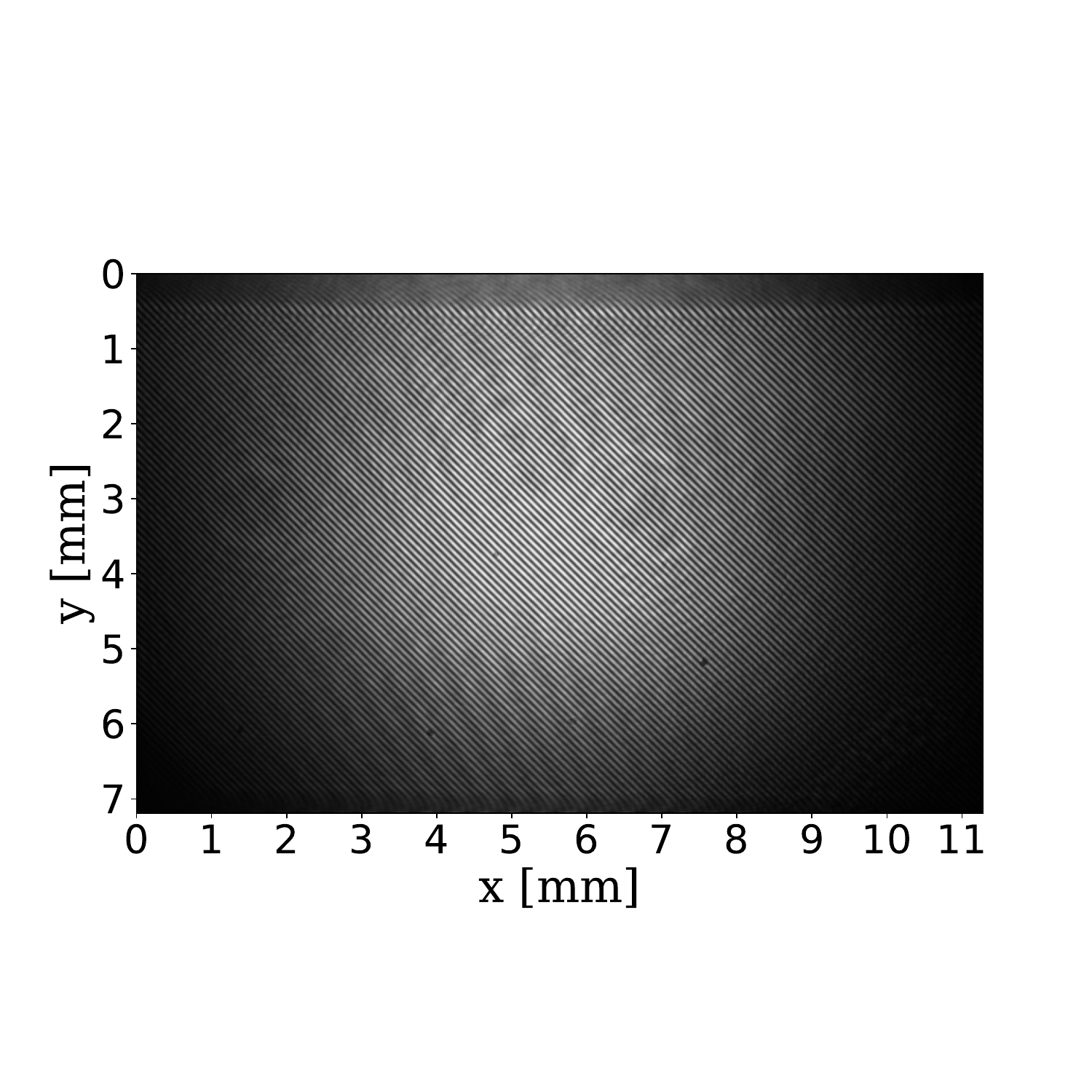}
    \includegraphics[width=\columnwidth, clip, trim = 0cm 0cm 0cm 1cm]{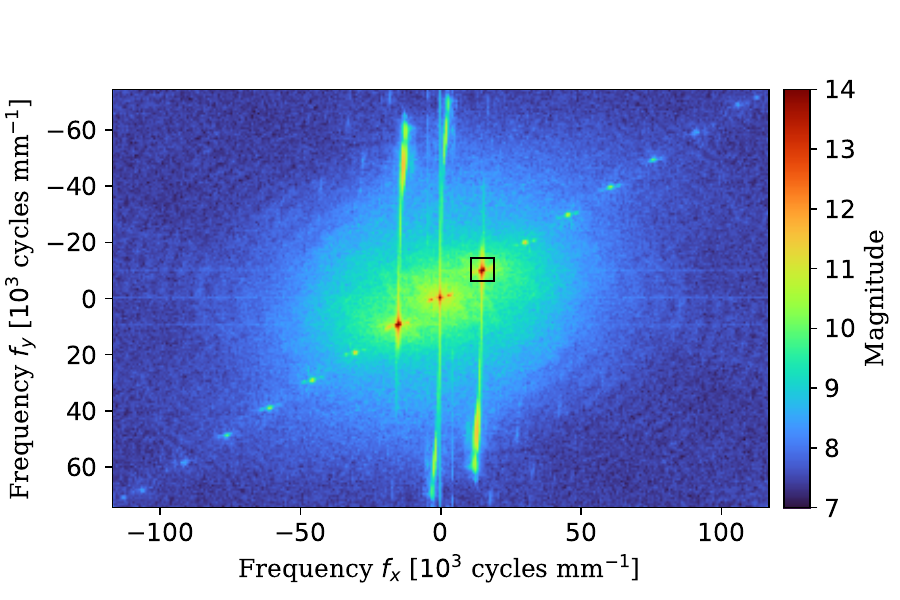}
    \caption{Top: A typical Mach-Zehnder interferogram obtained with a cylindrical nozzle at \qty{6}{bar} inlet pressure. A nozzle tip is visible as a dark shadow on the top of interferogram. -- Bottom: Fourier transform of the interferogram. The selected frequency band is indicated by a black square.}
    \label{fig:Interferogram}
    \label{fig:fourier_transform}
\end{figure}

For the interferometer data acquisition, we record 100 interferograms each for the cases with and without gas jet. 
We then compare pairs of interferograms with and without jet. Using an automated method, we pre-select about a dozen pairs where the background sidebands, i.e. the areas outside the jet region, match. The pre-selected pairs are then visually inspected for unphysical artefacts, and one pair with a regular background pattern is retained for the further analysis (\cref{fig:fourier_transform}, top panel). 

\label{sec:phase_retrieval}

The theory of Mach-Zehnder interferometry is well established \cite{Takeda82-JOptSocAm,Malka2000-RevSciInstrum,Couperus16-NIMA}. Here, we summarize the formalism used by us. 

In  the interferogram without gas, and also in the sidebands parts of the interferogram with gas, the intensity pattern is given by
\begin{equation}
    I_{\rm no-jet}(x,y) = a(x,y) + b(x,y) \left[ \cos\left(\phi(x,y\right) \right]
    \label{eq:interferogram}
\end{equation}
where $a(x,y)$ and $b(x,y)$ parameterize disturbances like air flow in the setup and phase shifts due to nonuniform light scattering, $x$ and $y$ are the spatial coordinates perpendicular to the direction of propagation of the light, and $\phi(x,y)$ is the position-dependent phase difference between the two arms of the interferometer. 

When the gas jet is on, an additional optical path length difference, here given as phase shift $ \Delta \phi(x,y)$, is introduced in the part of the interferogram where the jet image is seen:
\begin{equation}
    I_{\rm jet}(x,y) = a(x,y) + b(x,y) \left[ \cos\left(\phi(x,y) + \Delta \phi(x,y)\right) \right]
    \label{eq:jet_interferogram}
\end{equation}

As a next step, each of the two interferograms $I_{\rm jet}(x,y)$ and $I_{\rm no-jet}(x,y)$ is Fourier transformed with respect to the $x$ direction. The resulting Fourier spectrum (\cref{fig:fourier_transform}, bottom panel) contains the zero spatial frequency at the center and two side frequencies. The Fourier transformation of the interferogram with jet can be expressed as
\begin{equation}
    \mathcal{F}\{I_{\rm jet}(x,y)\} = A(x,y) + C(f-f_0,y)  + C^{*}(f + f_0, y) ]
    \label{eq:fourier_transform}
\end{equation}
where the capital letters $A$, $C$, and $C^*$ represent the Fourier spectra and $f$ is the spatial frequency in $x$ direction.

For further analysis, one of the two side frequencies in the Fourier domain is selected (black box in \cref{fig:fourier_transform}). This side frequency data is then shifted by $f_0$ towards the origin to filter out noise components outside the selected region. The filtered component is then simply given by:
\begin{equation}
    \left.\mathcal{F}\{I_{\text{jet}}(x,y)\}\right|_{f_1} = C(f - f_0, y)
\end{equation}
with the components $A$ and $C^*$ set to zero.

Subsequently, the inverse Fourier transformations of the selected and shifted side frequencies were computed. The real part contains background variations, while the imaginary part contains the phase information needed for the further analysis. For the jet and no-jet interferograms, respectively, the following phase maps were found:
\begin{align*}
    \phi(x,y) + \Delta \phi(x,y) 
    = & \arctan\frac{\operatorname{Im}\left[ \mathcal{F}^{-1}\left\{\left.\mathcal{F}\{I_\text{jet}(x, y)\}\right|_{f_1}\right\} \right]}{\operatorname{Re}\left[ \mathcal{F}^{-1}\left\{\left.\mathcal{F}\{I_\text{jet}(x, y)\}\right|_{f_1}\right\} \right]} \\ 
    \phi(x,y) 
    = & \arctan\frac{\operatorname{Im}\left[ \mathcal{F}^{-1}\left\{\left.\mathcal{F}\{I_\text{no-jet}(x, y)\}\right|_{f_1}\right\} \right]}{\operatorname{Re}\left[ \mathcal{F}^{-1}\left\{\left.\mathcal{F}\{I_\text{no-jet}(x, y)\}\right|_{f_1}\right\} \right]} 
\end{align*}
The final interferogram was then found by subtracting the no-jet from the jet phasemap, resulting in the phase shift $\Delta\phi(x,y)$ due to the jet.

\begin{figure*}[tb]
    \centering
    \includegraphics[width=\textwidth]{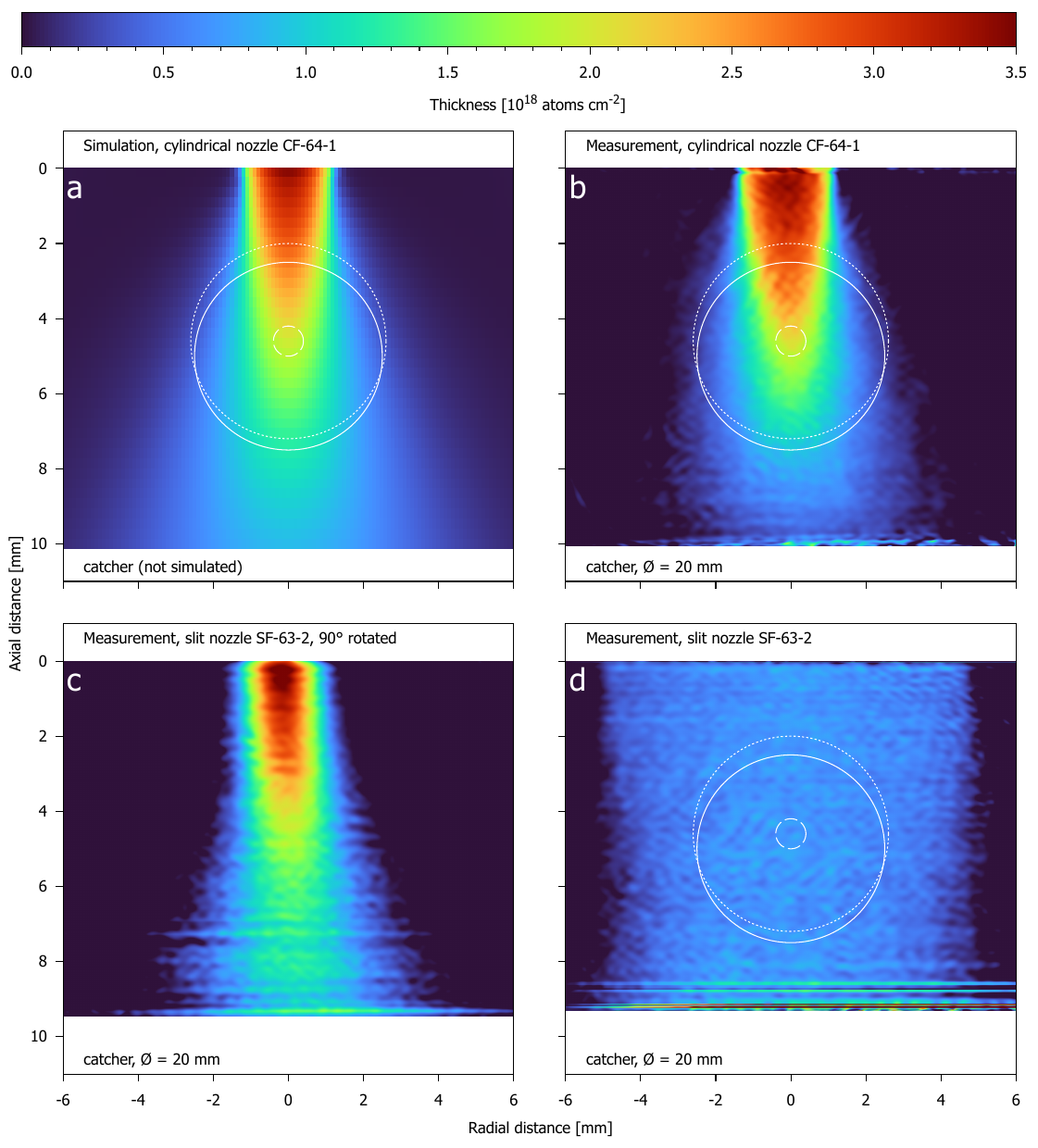}
    \caption{\label{fig:sim_and_int} Simulated (a) and interferometrically measured (b--d) thickness profiles of the nitrogen jet from nozzles CF-64-1 (a, b) and SF-63-2 (c, d) at an inlet pressure of \qty{6}{bar}. See text for details. The solid white circle denotes the region relevant to the ion-beam passage: a circle of \qty{5.0}{mm} diameter centered \qty{5.0}{mm} from the nozzle. The dotted and dashed circles show the regions where the $\alpha$ energy loss measurements were done with uncollimated ($d$=\qty{5.2}{mm}, \qty{4.6}{mm} from the nozzle) or collimated ($d$=\qty{0.8}{mm}, \qty{4.6}{mm} from the nozzle) Si detector, respectively, and an $\alpha$ source. The blurring observed beneath the nozzle and above the catcher is due to reflections off these nearby surfaces.%
    }
\end{figure*}


As next step, the phase shift $\phi$ was converted to gas thickness $t$ in units of cm$^{-2}$, using
\begin{eqnarray}\label{eq:Thickness}
    t(x,y) & = &  \frac{\rho}{A_r u} \, \frac{\lambda}{2 \pi (n-1)} \, \Delta\phi(x,y) \nonumber \\
    & = & 1.815\times10^{18}\,  {\rm cm}^{-2} \, \Delta\phi(x,y)
\end{eqnarray}
with the laser wavelength $\lambda$ = \qty{632.8}{\nm}, refractive index $n$ = 1.0002751 for this wavelength \cite{Rumble2024}, $\rho$ = 1153 g/m$^3$ the mass density, $A_r=14.0067$ the relative atomic weight of the nitrogen atom, and $u$ = \qty{1.6605e-24}{g} the atomic mass unit. The values for $n$ and $\rho$ have been taken for nitrogen gas at 23$^\circ$C room temperature and normal pressure, 1013.25 hPa. 

Applying the conversion factor (\ref{eq:Thickness}) to the interferometric map, an experimental jet thickness map is obtained (\cref{fig:sim_and_int} b -- d).

\section{Energy loss measurements with an alpha source}
\label{sec:AlphaEnergyLoss}
\label{subsec:AlphaEnergyLoss}

\begin{figure}
    \centering
	\begin{annotationimage}{width=\columnwidth}{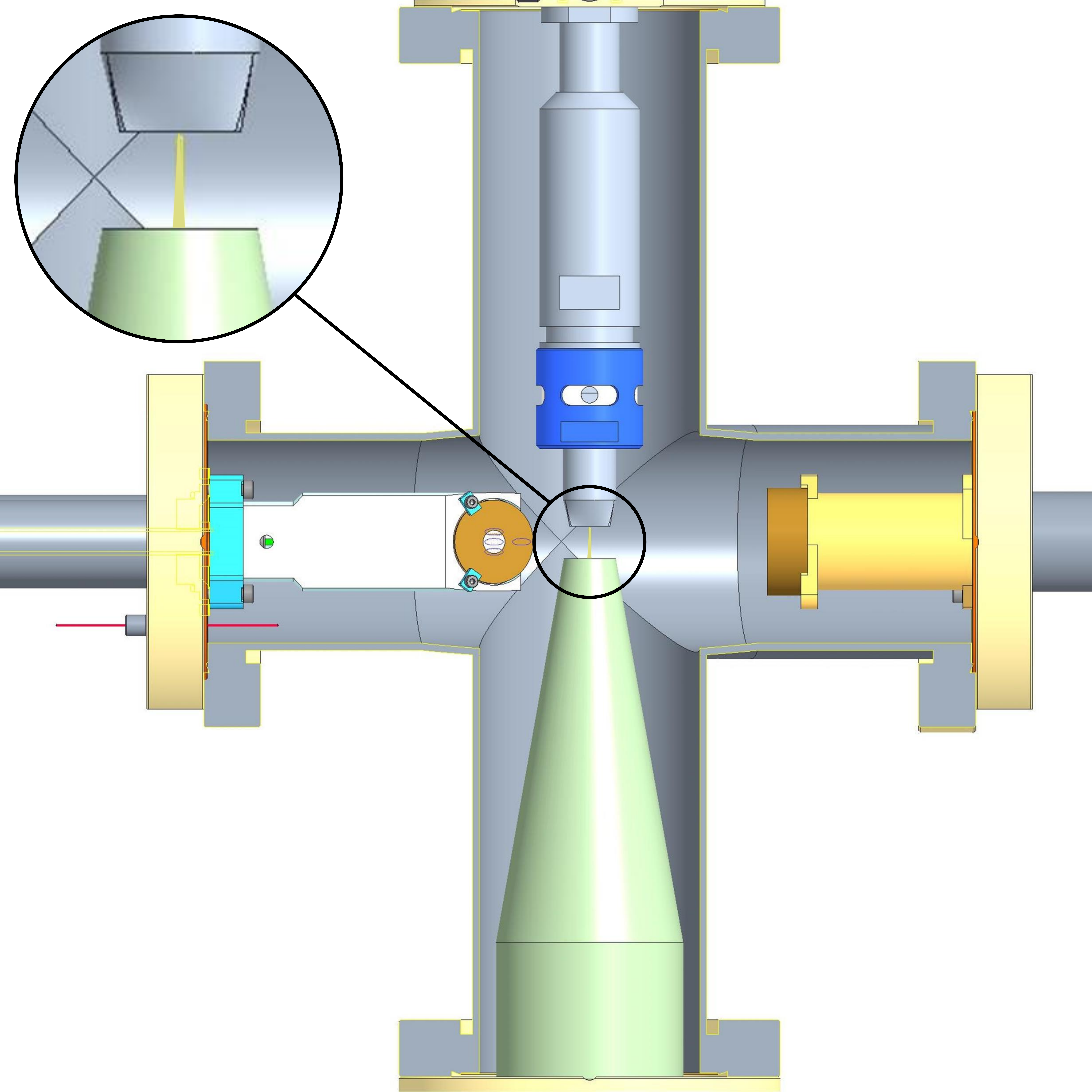}
		\imagelabelset{
			coordinate label font = \normalfont\small,
			coordinate label text = black,
			coordinate label style/.style = {
				text = \coordinatelabeltext,
				font = \coordinatelabelfont,
			}
		}
		\draw[coordinate label = {jet 				at (0.210,0.835)}];
		\draw[coordinate label = {nozzle 			at (0.540,0.800)}];
		\draw[coordinate label = {silicon 			at (0.805,0.525)}];
		\draw[coordinate label = {detector 			at (0.815,0.485)}];
		\draw[coordinate label = {$\alpha$-source 	at (0.335,0.500)}];
		\draw[coordinate label = {catcher 			at (0.540,0.200)}];
	\end{annotationimage}
    \caption{Experimental setup for the $\alpha$-particle energy loss measurement, including the gas jet, the $\alpha$ source, and the particle detector. See text for details.}
    \label{fig:EnergylossSetup}
\end{figure}

\begin{figure*}[htb]
    \centering
    \includegraphics[width=\textwidth, clip, trim={2cm 0cm 0 0},clip]{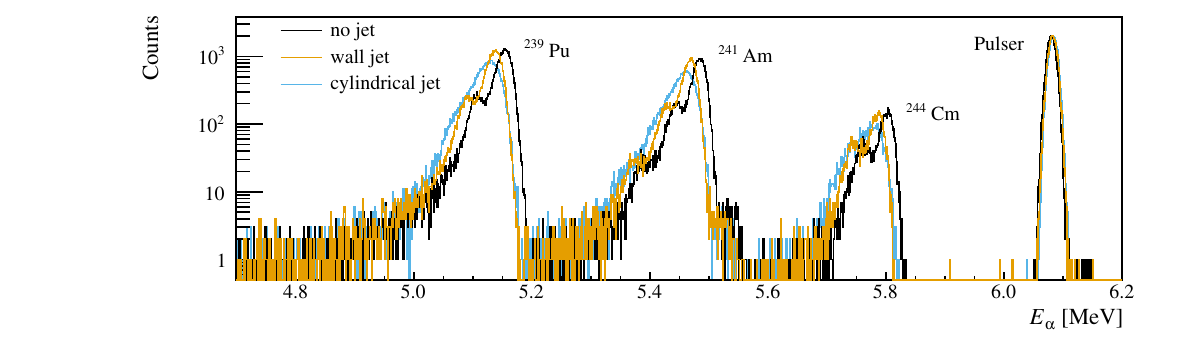}
    \caption{Typical spectra in the particle detector: Reference spectrum without gas (black), spectrum with slit-type nozzle and 0.77$\times$10$^{18}$ cm$^{-2}$ thickness (orange), spectrum with cylindrical nozzle and  1.49$\times$10$^{18}$ cm$^{-2}$ thickness (light blue). See text for details.}
    \label{fig:Alphaspectra}
\end{figure*}

In order to independently measure the thickness of the gas jet, $\alpha$-energy loss measurements were performed.

To this end, a mixed $\alpha$ source (\ce{^239Pu}, \ce{^241Am} and \ce{^244Cm}) was placed at an angle of 15$^\circ$ with respect to the wall jet normal opposite a \qty{300}{\square\mm} PIPS detector (model A300-17AM, nominal \qty{17}{\keV} resolution for the \ce{^241Am} peak at \qty{5.486}{\MeV}), see \cref{fig:EnergylossSetup} for the setup. A pulser signal was sent to the preamplifier test input to ensure gain stability and to measure the data acquisition dead time.

Typical particle detector spectra are shown in \cref{fig:Alphaspectra}. As expected, for the measurements with jet, the peaks shift to lower energies. In addition, they grow wider due to the structure of the gas jet, which is not uniform in the area irradiated by the $\alpha$-source (\cref{fig:sim_and_int}). 
When comparing the cylindrical and slit-type nozzles, it is found that the cylindrical nozzle shows larger energy shift and larger peak widening   (\cref{fig:sim_and_int}). 

The quantitative analysis is limited to the main peaks of the three isotopes in the mixed $\alpha$ source, corresponding to the most probable energy loss experienced by the $\alpha$ particles. To this end, a Gaussian is fitted in several iterations to the right half of each of the source peaks to determine the peak center. Typical energy loss values for the \ce{^239Pu} peak are \qty{21.2+-0.9}{\keV} for the cylindrical-type nozzle CF-64-1 and \qty{13.6+-0.9}{\keV} for the slit-type nozzle SF-63-2.  Using the SRIM stopping power \cite{Ziegler2010}, each peak shift is then converted to a gas thickness, and the weighted average of the three values is adopted. 

The $\alpha$ energy loss measurements have been carried out at several inlet pressures for nine different nozzles. In all cases, a linear dependence of target thickness on the inlet pressure is found. 
Furthermore, cylindrical nozzles were typically found to show greater jet thickness, as expected.

The highest thickness was measured at \qty{6}{bar}, which is the highest inlet pressure explored here. For this pressure, the results of all the $\alpha$ energy loss measurements are listed in \cref{tab:JetCharacteristics}. When taking the values without $\alpha$-collimator, the maximum thickness for a cylindrical nozzle is \qty{14.9+-0.8e17}{atoms/\square\cm}, and for the wall jet,  \qty{7.8+-0.5e17}{atoms/\square\cm}. The main sources of uncertainty in the $\alpha$ energy loss based target thickness are the stopping power value \cite{Ziegler2010} and the determination of the peak centroid from the fit.

\begin{table*}
    \centering
    \begin{tabular}{@{}
                    l
                    S[table-format=-1.3]
                    S[table-format=-1.2+-1.2]
                    S[table-format=-1.3]
                    S[table-format=-1.3]
                    S[table-format=-1.2+-1.2]
                    S[table-format=-1.3]
                    S[table-format=-1.3]
                    S[table-format=-1.3]
                    @{}}
        \hline
        \hline
        & \multicolumn{3}{c}{With $\alpha$-collimator (\qty{0.8}{\mm})} & \multicolumn{3}{c}{Without $\alpha$-collimator (\qty{5.2}{\mm})} & \multicolumn{2}{c}{Ion-beam size (\qty{5.0}{\mm})}\\
        \cline{2-4}\cline{8-9}
                & {Simu-}   &            {Energy}               & {Inter-}      & {Simu-}   &          {Energy} & {Inter-}      & {Simu-}                   & {Inter-}\\
        Nozzle  & {lation}  &  {loss} & {ferometry}     & {lation}  &  {loss}  & {ferometry}   & {lation}& {ferometry}\\
                \hline
        SF-61-2 &           &              & &           & 0.78+-0.05    & &           & \\
        CF-62-1 & 1.901     &              &1.96& 1.266     & 1.31+-0.05    &1.24& 1.254     &1.22\\
        SF-63-2 &           & 0.77+-0.11$^1$    & 0.76  &           & 0.77+-0.05    & 0.72  &           & 0.72 \\
        CF-64-1 & 1.935     & 1.98+-0.15$^2$    &1.93& 1.289     & 1.36+-0.09    &1.14& 1.273     &1.11\\
        CH-65-1 & 1.954     &              &1.98& 1.303     & 1.49+-0.08    &1.26& 1.282     &1.26\\
        CH-65-2 & 1.926     &              &1.98& 1.284     & 1.30+-0.07    &1.26& 1.263     &1.24\\
        CH-65-3 & 1.916     &              &1.92& 1.276     & 1.32+-0.07    &1.21& 1.255     &1.18\\
        CH-65-4 & 1.907     &              &2.05& 1.270     & 1.20+-0.07    &1.34& 1.249     &1.32\\
        CH-65-5 & 1.877     &              &1.80& 1.246     & 0.97+-0.06    &1.06& 1.226     &1.04\\
        \hline
        \hline
    \end{tabular}
    \footnotesize\ $^1$ collimated energy loss of nozzle SF-63-2 was studied at 5 bar instead of 6 bar inlet pressure. --- $^2$ This energy loss value is only based on the $^{239}$Pu peak shift.
    \caption{\label{tab:JetCharacteristics} Jet gas target thickness, in units of 10$^{18}$\,cm$^{-2}$, for a nitrogen jet with \qty{6}{bar} inlet pressure for various nozzles (dimensions in \cref{tab:NozzleDimensions}). Thickness data from the simulation, from the $\alpha$-energy loss, and from interferometry are shown for 0.8 and 5.2 mm diameter $\alpha$-collimators. The final two columns show simulated and interferometric thicknesses over the approximate area of the ion beam (5\,mm diameter). See text for details and \cref{fig:Consistency} for an illustration. 
    }
\end{table*}

\begin{figure*}[bt]
    \includegraphics[width=\textwidth, clip, trim={0cm 0cm 0cm 0cm}]{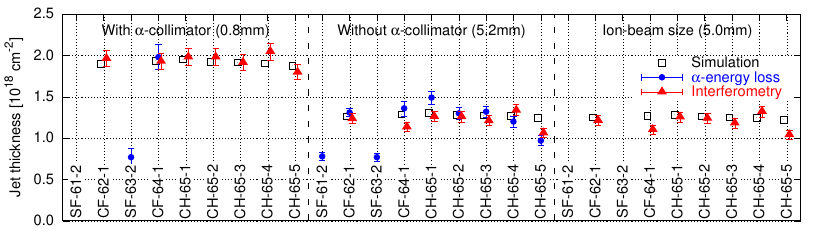}
    \caption{Jet gas thickness for nitrogen jet, 6 bar inlet pressures, for different nozzles and scenarios. See Table \ref{tab:JetCharacteristics} for the numerical data. An uncertainty of 5\% is assumed for the interferometric values.}
    \label{fig:Consistency}
\end{figure*}

\section{Discussion}
\label{sec:Discussion}

For the cylindrical-type nozzles, excellent qualitative agreement is found between the one-dimensional computational fluid dynamics simulation and the interferometric image of the jet (\cref{fig:sim_and_int} a and b). In both simulation and interferometry, the map shows how the jet expands sideways with a strong thickness gradient downwards.

The qualitative agreement between simulation and interferometry shows that for the cylindrical symmetric case, the simulation gives a reasonable description of the situation. In particular, the two main approximations seem to be justified, namely to neglect the outer walls of the target chamber and the gas catcher. 

The simulation does not include the catcher. As a result, the simulated jet keeps diverging with the same slope over the whole distance between nozzle and the lower end of the simulated area. In the experiment, a catcher is placed \qty{10}{\mm} downstream from the nozzle, capturing most of the gas and contributing to a more drop-like shape of the jet (\cref{fig:sim_and_int}). When comparing the lower left and right corners of simulated and experimental images, the effects of the catcher are visible. Therefore, these corners are excluded from the determination of the background area for matching experimental jet and no-jet interferograms.  

When studying the slit-type nozzle, there is a marked difference in the density maps depending on the angle of observation (\cref{fig:sim_and_int} c and d). While the view from the side shows a $<$2\,mm thick gas curtain, the view perpendicular to the wall shows a uniform, 10\,mm wide wall, with a much lower average thickness because the same amount of gas is spread out over a much larger area.

In \cref{tab:JetCharacteristics} and \ref{fig:Consistency}, the measured thickness values of nine different cylindrical nozzles are compared with the simulated thickness, for several scenarios. It is found that the simulation agrees with the measured thickness, when assuming a relative uncertainty of 5\% for the interferometric value. For those cases where $\alpha$ energy loss data are available, these data confirm what had been found by interferometry and by simulations. Due to the fact that only cylindrically symmetric jets were simulated here, no simulations are available for the wall jet created by slit-type nozzles.

The are two main types of measurements with the $\alpha$-energy loss method: The ones with the small $\alpha$-collimator, with an effective diameter of 0.8\,mm taking into account the possible paths of the $\alpha$ particles, probe the center of the jet. On the other hand, the measurements with the 5.2\,mm effective diameter collimator probe a representative average of the jet. As a result, for the slit-type nozzle which creates a 10\,mm wide gas curtain (\cref{fig:sim_and_int} c, d), the measured average thickness values are the same for the two collimators (\cref{tab:JetCharacteristics}). As expected, this is not the case for the cylindrical type nozzle, due to the much stronger structure of the jet (\cref{fig:sim_and_int} b).

When quantitatively comparing the thicknesses from the three methods, two out of the nine nozzles are outliers, both of them cylindrical. First, for the nozzle CF-64-1, the respective interferometric thickness value is 12-14\% below the $\alpha$-energy loss and simulation results for the scenarios without an $\alpha$-collimator. However, for the scenario with a very narrow $\alpha$-collimator, good agreement is found for this nozzle. This may point to some problems with the background subtraction in the interferometry in this particular case. Any overestimation of the background would lead to an underestimation of the total jet thickness especially for the runs with a wide averaging area, i.e. without $\alpha$-collimator or with the ion beam. This underlines the need to carefully measure the background in the sidebands not affected by the jet, to ensure a proper subtraction. 

The second outlier is nozzle CH-65-5, which has an extraordinarily long throat (4\,mm, \cref{tab:NozzleDimensions}). There, the interferometry result is 15\% lower than the simulation. For this nozzle, $\alpha$-energy loss data is available and confirms the interferometry. It is possible that the long throat somehow affected the simulation, leading to an overestimation of the simulated jet thickness. 

For the seven other nozzles studied, simulation, interferometry, and $\alpha$-energy loss are in excellent agreement (\cref{fig:Consistency}). This confirms the applicability of the interferometric method.

The average pressure in the jet is 37\,hPa, corresponding to 1.5$\times$10$^{18}$\,cm$^{-2}$ averaged over 5\,mm jet depth, for the most efficient nozzle (\cref{tab:JetCharacteristics}). This is $\sim$300 times larger than the ambient pressure in the jet chamber surrounding it (\cref{fig:PressureProfile}), showing very good contrast of the present system, with only negligible energy loss in the parts of the chamber before the jet. 

\section{Summary and outlook} 
\label{sec:Summary}
\label{sec:Outlook}

\subsection{Summary}

A jet gas target system for nuclear astrophysics has been developed, tested, and characterized using nitrogen gas. 

The target system consisted of altogether five pumping stages separating a room-temperature jet of up to 1.49$\times$10$^{18}$\,cm$^{-2}$ from the accelerator vacuum of better than 10$^{-7}$\,hPa. It has been shown that the jet thickness can be tuned using its proportionality to the inlet pressure, by varying inlet pressures from 0-6 bar. 

A beam calorimeter with constant temperature gradient for the determination of the beam intensity has been tested for hot side temperatures of 50-100\,$^\circ$C, giving a typical no-beam power of 100\,W at 80\,$^\circ$C hot side temperature. The vibrations caused by the pumps of the jet gas target system have been measured with accelerometers and were mitigated using a mechanical separation of several parts of the setup and the appropriate type of vibration attenuating supports, especially for the Okta Roots pump. 

The jet thickness has been determined consistently using three different methods, namely a computational fluid dynamics simulation assuming rotational symmetry, a Mach-Zehnder interferometer, and the $\alpha$-energy loss method. In particular, the interferometric determination of the target thickness has been shown to be consistent with the other two methods within 5\% uncertainty, provided that the  background in the interferometric image is determined with the requisite care. 

Various combinations of nozzles and catchers have been tested to determine the optimal setup configuration. A number of cylindrical and slit-type nozzles were tested, both made from stainless steel and from glass, and were shown to gave satisfactory results as regards the total jet thickness.

\subsection{Outlook}

All the development work and tests described here had been performed at the surface of the Earth, at HZDR's overground Rossendorf campus, and without incident ion beam. 

The system has in the mean time been transported to the underground Felsenkeller site \cite{Bemmerer25-EPJA} and successfully commissioned there using an $\alpha$-beam on a nitrogen jet gas target. Those data are now under analysis and will be reported in forthcoming work. 

Further in the future, the already installed windowless static-type gas target part behind the jet gas target chamber will be commissioned and characterized, both without and with incident ion beam. 

\section*{Acknowledgements}
Financial support by the European Union (ChETEC-INFRA, 101008324) is gratefully acknowledged. 


\end{document}